\documentclass[aps,prb,reprint,amssymb,superscriptaddress]{revtex4-2}

\usepackage{xcolor}

\usepackage{bm}

\usepackage{amsmath}  
\usepackage{amsfonts}
\usepackage{graphicx}
\usepackage{hyperref}
\hypersetup{
	colorlinks=true,
	citecolor=blue,
	linkcolor=red,
	urlcolor = blue
}

\begin{document}
	
\title{Benchmarking the generalized Kadanoff-Baym ansatz and second-order adiabatic expansion using time-dependent spintronic effects: Spin pumping, torque, and inertia}
    
\author{Jalil Varela-Manjarres}
\affiliation{Department of Physics and Astronomy, University of Delaware, Newark, DE 19716, USA}
\author{Nicole Sofia Salazar-Delgado}
\affiliation{Department of Physics and Astronomy, University of Delaware, Newark, DE 19716, USA}
\affiliation{Department of Physics, Universidad Nacional de Colombia, AA 14490 Bogot\'a, Colombia}
\author{Branislav K. Nikoli\'c}
\email{bnikolic@udel.edu}
\affiliation{Department of Physics and Astronomy, University of Delaware, Newark, DE 19716, USA}
	
%	\date{\today}
		
\begin{abstract}
The generalized Kadanoff-Baym ansatz (GKBA) [P. Lipavsk\'{y} {\em et al.}, Phys. Rev. B {\bf 34}, 6933 (1986)] has emerged as a popular and  numerically efficient algorithm for simplification of   nonequilibrium Green's function (NEGF)-based calculations of time-dependent quantum transport. This is because it offers linear scaling of total computational time in the number of time steps. Algorithms with such scaling are, in turn, crucial to reach relevant  $\sim 1$~ps or $\sim 1$~ns time scales of phenomena in nanoelectronics and spintronics. For systems that can be split into classical and quantum degrees of freedom, another popular simplifying strategy is adiabatic expansion (AE) of NEGF [N. Bode {\em et al.}, Phys. Rev. Lett. {\bf 107}, 036804 (2011); S. Deghi {\em et al.}, Phys. Rev. B {\bf 110}, 115409 (2024)] in terms of the velocity of classical degrees of freedom, such as localized magnetic moments (LMMs) in spintronics or coordinates of nuclei in nanoelectronics. Here we compare GKBA and second-order AE with numerically exact benchmarks for two-terminal junctions whose central region hosting quantum electrons and classical LMMs is attached to two semi-infinite normal metal leads. Three simple models are employed to exhibit  cornerstone time-dependent effects in spintronics---spin pumping and spin-transfer torque (STT), as well as magnetic inertia as a recently explored phenomenon. We find that GKBA fails to describe pumping of spin current by precessing LMMs, or STT vectors, and thereby induced LMM dynamics. Conversely, the second-order AE matches numerically exact benchmarks for both effects remarkably well, thereby also revealing the essentially {\em nonadiabatic} nature of spin pumping. Thus, AE opens a path toward an accurate description of STT-driven magnetization dynamics, including combination with first-principles Hamiltonians, while incurring a fraction of the cost of time evolution of full NEGF. However, despite including terms up to the second time derivatives of LMMs into AE, this approach fails to capture fast nutational oscillations on top of the precessional motion of LMMs as the hallmark of magnetic inertia. 
\end{abstract}

\maketitle

\section{Introduction}\label{sec:intro}

The nonequilibrium Green's function (NEGF)~\cite{Stefanucci2025,Schlunzen2019,Gelis2019} formalism is presently the only approach capable of studying the dynamics of quantum 
many-body systems for long times and in two and three dimensions, for both closed~\cite{Schlunzen2019,Gelis2019} and  open~\cite{Sieberer2016,Stefanucci2024,ReyesOsorio2026} (due to coupling to a dissipative  environment) systems.   It is not limited to weak driving or linear response only;  many-particle correlations can be systematically included~\cite{Stefanucci2025,Schlunzen2019,Gelis2019}  by diagrammatic construction of self-energies; and  dependence of its Green's functions (GFs) on two different times, $t$ and $t'$,  provides access to both time-dependent observables and their correlators~\cite{ReyesOsorio2026}. For example, evolving the density  matrix $\hat \rho(t)$, which is a function of single time, provides access to expectation values  $\langle \hat O \rangle(t)=\mathrm{Tr}[\hat{\rho}(t) \hat{O}]$ for  any observable $\hat O$, but not to their correlators~\cite{ReyesOsorio2026} at different times~\cite{Fux2024}.  

\begin{figure}
		\centering
		\includegraphics[scale=0.3  ]{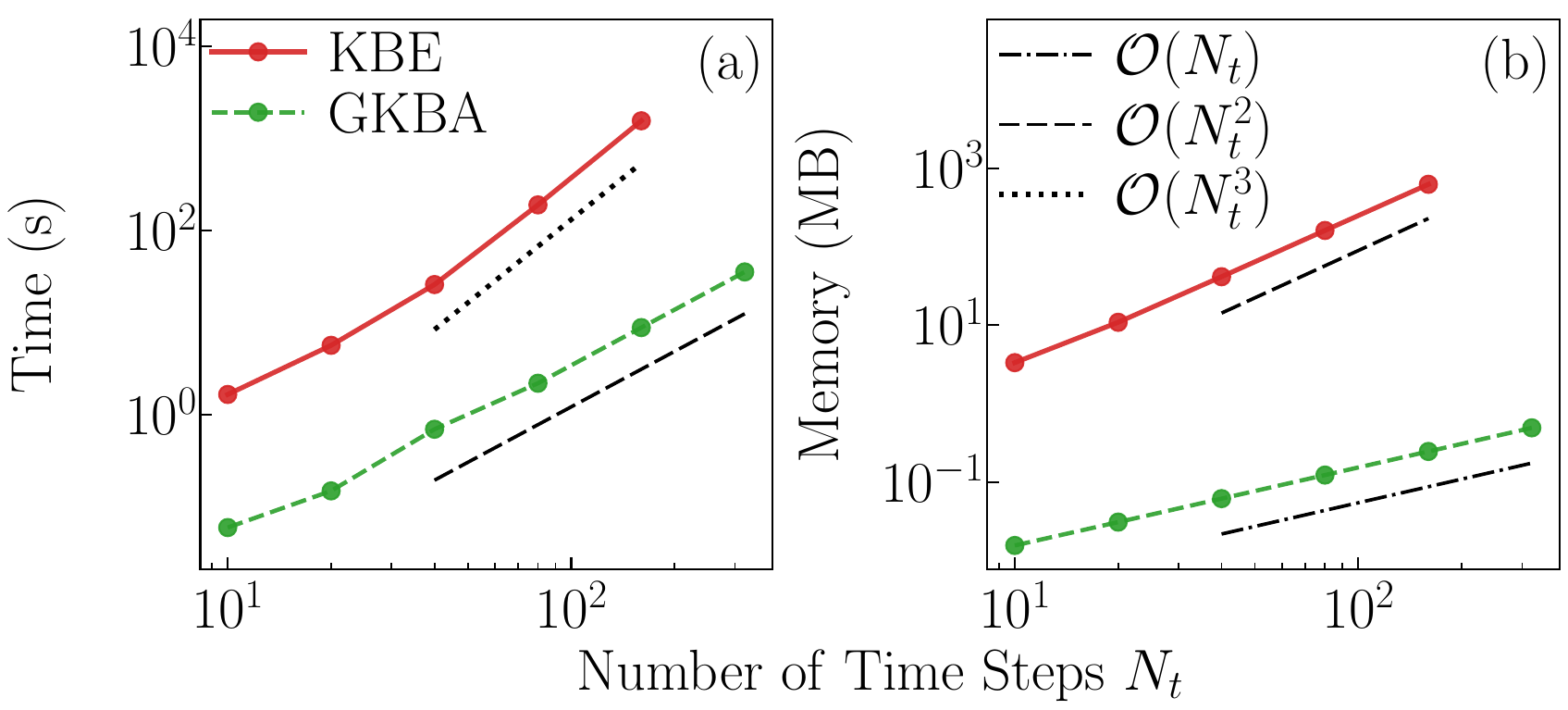}
		\caption{
        Scaling of (a) total computational time and (b) computer memory with a number of time steps $N_t$ in numerical algorithms solving the full KBE~\cite{Stefanucci2025} or  GKBA~\cite{Stefanucci2025,Lipavsky1986} applied [Fig.~\ref{fig:fig2}] to spin pumping setup of Fig.~\ref{fig:fig0}(a).  For easy comparison, linear, quadratic and cubic scaling are plotted as   dash-dotted, dashed and dotted black straight  lines, respectively. 
    }
    \label{fig:time_complexity}
\end{figure}

However, NEGFs depending on two times and time-convolutions within their equations of motion, known as the Kadanoff-Baym equations (KBE)~\cite{Stefanucci2025,Schlunzen2019,Gelis2019}, make brute-force time-stepping numerical algorithm scaling [Fig.~\ref{fig:time_complexity}(a)] at least cubically with the number of time steps $N_t$, i.e., acquiring computational complexity~\cite{Mertens2002} $\mathcal{O}(N_t^3)$. Thus, for long evolution times, numerical simulations become prohibitively expensive. They also require large amounts [Fig.~\ref{fig:time_complexity}(b)] of computer memory~\cite{Kaye2021,Sroda2025}.  This means that realistic time-dependent quantum transport phenomena encountered in  nanoelectronics, nanophononics,  spintronics,  and magnonics are apparently unreachable by such simulations, as the relevant experimental time scales \mbox{$\lesssim 1$ ps} would take too much time (using the typical time step \mbox{$\delta t \lesssim 0.1$ fs} required for numerical stability~\cite{Gaury2014,Kloss2021,Popescu2016,Pavlyukh2023,Petrovic2018}) or excessive amounts of memory. This obstacle has prompted various algorithmic improvements~\cite{Kaye2021,Meirinhos2022,Blommel2024,Murray2024,Sroda2025,Reeves2024,Reeves2025,Lamic2025} for solving the full two-time integro-differential KBE [Sec.~\ref{sec:kbe}], making it possible to reduce the computational complexity from $\mathcal{O}(N_t^3)$ to $O(N_t^2 \log N)$, as well as the memory complexity from $\mathcal{O}(N_t^2)$ to $O(N_t \log N)$~\cite{Golez2014,Sroda2025}.  Nevertheless, these advances remain short of the desired {\em linear-scaling} computational complexity $\mathcal{O}(N_t)$. Extrapolation techniques have also been explored by   learning dynamics from short time samples of data~\cite{Yin2022}, including neural network-based nonlinear operator learning~\cite{Zhu2025}.

A popular alternative to improving the efficiency of solving the KBE is to make controlled approximations to them. Among such approximations, the so-called generalized KB ansatz (GKBA) has attracted continuous attention since its introduction in 1986~\cite{Lipavsky1986}. The original GKBA exhibits $ \mathcal{O}(N_t^2)$ computational complexity [green curve in Fig.~\ref{fig:time_complexity}(a)]. Furthermore,  recent efforts~\cite{Bonitz2024,Karlsson2021}  modifying the original GKBA  have achieved the desired linear scaling, for  either bulk systems or multiterminal junctions, but typically using simplified self-energies of the semi-infinite leads in the so-called  wide band limit (WBL) leads~\cite{Latini2014,Bruch2016} in the latter case. Thus, GKBA-WBL exhibiting linear scaling has become a popular foundation for coding   packages for computational {\em time-dependent} quantum transport of electrons~\cite{Pavlyukh2023,Tuovinen2023} interacting  with each other or with bosonic quasiparticles~\cite{Karlsson2021} in multiterminal junctions. Other linear-scaling packages---such as {\tt TKWANT}~\cite{Gaury2014,Kloss2021},  {\tt Zandpack}~\cite{BachLorentzen2026} or {\tt TDNEGF+LLG}~\cite{Petrovic2018,Bajpai2019a,qttgsoftware}---are designed {\em  only} for the transport of noninteracting electrons and are accelerated by using vectors instead of matrix representations of NEGFs, so they do not allow for the inclusion of many-body interactions (via their respective self-energies~\cite{Stefanucci2025,Schlunzen2019,Gelis2019}). 

The GKBA effectively neglects~\cite{Reeves2023,Reeves2024} certain memory effects that otherwise contribute to NEGFs at $t\neq t'$. Although this strategy can be highly successful in situations where neglected terms are much smaller than those retained, their ratio is not known in advance. This issue has prompted numerous studies performing benchmarking~\cite{Kalvova2024,Tuovinen2020} of GKBA vs. the full KBE. For example, recent such benchmarking for the cases of bulk excitonic insulator~\cite{Tuovinen2020} or Hubbard clusters with sufficiently strong on-site Coulomb interaction~\cite{Hopjan2026} finds significant discrepancies. Other comparisons of GKBA vs. full KBE have been focused on charge transport in two-terminal junctions, often finding  an excellent  agreement~\cite{Cosco2024,Latini2014,Pavlyukh2025}. However, they have typically been focused  on reproducing {\em steady-state} charge current in the long time limit, as driven by turning on a DC bias voltage, thereby {\em lacking} test cases where genuinely time-dependent quantum transport phenomena or their dependence on spin is exhibited. 

\begin{figure}
		\centering
		\includegraphics[scale=1.9]{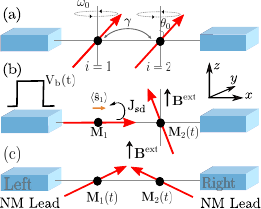}
		\caption{Schematic view of 1D toy models~\cite{Ralph2008,Chen2009} of two-terminal junctions exhibiting cornerstone effects of spintronics: (a) pumping of spin currents~\cite{Tserkovnyak2005,Ando2014a,VarelaManjarres2023} into the $L$ and $R$ NM leads due to two LMMs (red vectors) steadily precessing with frequency $\omega_0$ and cone angle $\theta_0$ in the \textit{absence} of any bias voltage; (b) voltage pulse-driven  STT~\cite{Ralph2008} where fixed LMM at site $i=1$ polarizes spin (orange arrow) of electrons of injected current, which then exerts torque on free-to-move LMM at site $i=2$ causing  its time evolution $\mathbf{M}_2(t)$ in accord with the  LLG equation~\cite{Landau1935,Evans2014,Berkov2008}; and (c) magnetic inertia~\cite{Mondal2023} manifesting  as fast nutational oscillations superimposed on top of precessional motion of two LMMs whose dynamics is initiated by the applied magnetic field along the $z$-axis. In all three junctions,  two TB sites (black circles) and two LMMs they host serve as the central region that is attached to two ideal semi-infinite NM leads modeled by TB  chains terminating at infinity into macroscopic electronic reservoirs. The reservoirs are kept at the same chemical potential in (a) and (c), or we introduce a voltage pulse through the $L$ lead in (b).}
        \label{fig:fig0}
\end{figure}

In this study, we perform a comparison of GKBA or GKBA-WBL, tailored for systems with semi-infinite NM leads~\cite{Balzer2023,Tuovinen2023},  vs. the full KBE calculations for three models of  two-terminal spintronic junctions depicted in Fig.~\ref{fig:fig0}. These junctions are chosen to be both simple and to exhibit  cornerstone  {\em time- and spin-dependent} quantum transport effects in spintronics, such as spin  current pumping~\cite{Tserkovnyak2005}, spin transfer  torque (STT)~\cite{Ralph2008} and magnetic inertia~\cite{Mondal2023}. The junctions consist of a central (C) active region of two tight-binding (TB) sites between which conduction electrons hop while interacting with local magnetic moments (LMMs) that are treated as classical vectors (denoted by red arrows in Fig.~\ref{fig:fig0}) of unit length, $\mathbf{M}_i$. Since all three junctions in Fig.~\ref{fig:fig0} are also examples of quantum-classical systems, composed of a quantum subsystem of conduction electrons and a classical subsystem of LMMs, adiabatic expansion (AE)~\cite{Bode2011,Bode2012,Deghi2024,Xu2023b} of NEGFs  can also be employed as a computationally much faster alternative to solving the full KBE. Thus, we benchmark second-order AE~\cite{Deghi2024} as well by comparing it to the exact solutions for systems in Fig.~\ref{fig:fig0} obtained from the full KBE.

Spin pumping is a phenomenon~\cite{Tserkovnyak2005,Ando2014a,VarelaManjarres2023} where steadily precessing LMMs generate a spin current flowing out into two semi-infinite normal metal (NM) leads in Fig.~\ref{fig:fig0}(a) in the {\em absence} (hence ``pumping''~\cite{Citro2023} terminology) of any bias voltage applied between them. In STT~\cite{Ralph2008} phenomenon, typically~\cite{Baumgartner2017} a voltage pulse is applied, such as in the left NM lead in Fig.~\ref{fig:fig0}(b), to inject electrons that become spin-polarized in the direction of fixed LMM $\mathbf{M}_1$. Since this direction is noncollinear to the second LMM $\mathbf{M}_2$ that is free to move, this free LMM receives torque $\mathbf{T}_2$ initiating its current-driven dynamics. Such dynamics is captured by the Landau--Lifshitz-Gilbert (LLG) equation~\cite{Evans2014,Berkov2008}
\begin{eqnarray}\label{eq:llg}
	\partial_t \mathbf{M}_i & =\! & -g_0 \mathbf{M}_i \times \mathbf{B}^\mathrm{ext} + \alpha_G \mathbf{M}_i \times \partial_t \mathbf{M}_i  + \frac{g_0}{\mu_M}\mathbf{T}_i, 
\end{eqnarray}
where $i=2$ for free LMM in Fig.~\ref{fig:fig0}(b) and we do not consider $i=1$ as that LMM is fixed in position. Here $\mathbf{B}^\mathrm{ext}$ is an external magnetic field; $\alpha_G$ is the Gilbert damping~\cite{Evans2014};  $g_0$ is gyromagnetic ratio~\cite{Evans2014}; $\mu_M$ is the magnitude of LMMs; and  we use shorthand notation $\partial_t \equiv \partial/\partial t$. The STT vector is obtained from~\cite{Nikolic2018,Belashchenko2019,Ellis2017}
\begin{equation}\label{eq:stt}
\mathbf{T}_i (t) = J_{sd}   \langle \hat{\mathbf{s}}_{i}  \rangle (t)  \times \mathbf{M}_i(t), 
\end{equation}
 where the nonequilibrium spin density 
 \begin{equation}\label{eq:sden}
     \langle \hat{\mathbf{s}}_i \rangle (t) = \mathrm{Tr}_\mathrm{spin} [{\bm  \rho}(t) {\bm \sigma}], 
 \end{equation}
 is  computed by tracing the product of  the vector of the Pauli matrices, \mbox{${\bm \sigma}=({\bm \sigma}_x,{\bm \sigma}_y, {\bm \sigma}_z)$}, and the nonequilibrium density matrix ${\bm  \rho}(t)$ in the spin space only. The $sd$~\cite{Ralph2008,Cooper1967} (or Kondo~\cite{Tsvelik2017}) exchange interaction between the  spin of conduction electrons and LMMs is denoted by $J_{sd}$. The same Eq.~\eqref{eq:llg} applies to the third junction in Fig.~\ref{fig:fig0}(c), where the dynamics of two LMMs is initiated by an external magnetic field  $\mathbf{B}^\mathrm{ext} = B^\mathrm{ext}  \mathbf{e}_z$ that is noncollinear to both of them. The STT is operative here as well~\cite{Petrovic2021}, despite no current being injected from NM leads, because dynamics of one LMM will cause spin current pumping by it, which then causes $\mathbf{T}_i[\mathbf{M}_i(t)]$ on the other LMM, and vice versa, in a self-consistent fashion~\cite{Petrovic2018,Petrovic2021}. Note that in the setup of Fig.~\ref{fig:fig0}(b), $\mathbf{T}_i[V_b(t)]$ is a functional of an externally applied voltage pulse only as the position of $\mathbf{M}_1$ is fixed. Two LMMs in Fig.~\ref{fig:fig0}(c) can then exhibit magnetic inertia~\cite{Mondal2023} as an effect due to a second derivative $I \partial^2_t \mathbf{M}_i$ within the  LLG equation, which in our setup is effectively generated~\cite{Bajpai2019a} by $\mathbf{T}_i[\mathbf{M}_i(t)]$. When the inertial term is important, it manifests as fast nutational oscillation on top of regular damped precessional motion. The recent time-resolved pump-probe experiments have started to detect and measure the nutation frequency~\cite{Neeraj2020,De2025}. In bulk magnetic materials, it has been related to spin-orbit coupling effects~\cite{Mondal2017}, unquenched orbital angular momentum~\cite{Moussa2026}, or the effect of different fermionic~\cite{ReyesOsorio2025,Svingen2025} and bosonic~\cite{Quarenta2024} baths with whom localized classical spins can interact. 
 
 All three effects studied in junctions of Fig.~\ref{fig:fig0} involve {\em semi-infinite} NM leads. They allow us to inject current, or to drain currents pumped by LMM dynamics even when no bias voltage is applied, as in the case of junctions in Figs.~\ref{fig:fig0}(a) and ~\ref{fig:fig0}(c). Of crucial importance  for dissipation effects in quantum transport~\cite{Joao2025}, in the absence of explicit inelastic processes (such as electron-phonon interaction) in the central region, is that semi-infinite leads provide  a continuous energy spectrum. Thus, for the junctions in Fig.~\ref{fig:fig0}, we use the exact analytical expressions  for the  self-energy of their semi-infinite NM leads. In the case of GKBA applied to Fig.~\ref{fig:fig0}(a), we also examine the popular simplification~\cite{Latini2014,Bruch2016} of lead self-energy yielding the GKBA-WBL algorithm.

The AE  for quantum-classical hybrid systems,  where separation of time scales makes it possible to consider fast quantum degrees of freedom interacting with the slow classical ones, can be traced to the work of Berry and Robins~\cite{Berry1993}. Examples of such hybrid systems include fast quantum electrons and slow classical nuclei in problems in  physics, chemistry, and biology where nuclei are handled by classical molecular dynamics (MD)~\cite{Dou2017a,Dou2018,Agostini2019} solving systems of Newton second law equations; or fast quantum electrons and slow classical LMMs in spintronics where LMMs are handled by the classical micromagnetics~\cite{Moreels2026,Berkov2008} or atomistic spin dynamics (ASD)~\cite{Evans2014} solving systems  of LLG equations~\cite{Landau1935}. For closed quantum-classical hybrid systems, Berry and Robins~\cite{Berry1993} expanded the electronic density matrix in terms of the velocity of the classical degree of freedom. The equation of motion for the classical degrees of freedom then acquires backaction forces like ``geometric magnetism''~\cite{Berry1993,Campisi2012,Bajpai2020,Kolodrubetz2017g} as the first nonadiabatic correction that is linear in velocity of the classical degrees of freedom. Additional frictional (terminology used in MD) or damping (terminology used in magnetism~\cite{Evans2014}) forces linear in velocity  appear upon making the quantum system open by coupling it to a thermal bath~\cite{Campisi2012,Gaitan1998}  or to macroscopic reservoirs of particles~\cite{Thomas2012}. In other words, open systems acquire a continuous energy spectrum, as opposed to a discrete one considered in Refs.~\cite{Berry1993,Campisi2012,Kolodrubetz2017g}. Upon driving an open quantum subsystem  out of equilibrium, wind force (in the Newton second law) or STT (in the LLG equation) will also emerge~\cite{Bode2011,Thomas2012,Lu2012,Dou2017a,Hopjan2018,Bajpai2020}.  For such open systems, usage of AE as an approximation to their NEGFs can significantly accelerate NEGF+MD or NEGF+LLG calculations, as all NEGF-based expressions can 
be {\em precomputed} [Sec.~\ref{sec:ae}] {\em before} running the simulation.

The paper is organized as follows. In Sec.~\ref{sec:hamiltonians} we introduce Hamiltonian models of three junctions in Fig.~\ref{fig:fig0}, and then in Secs.~\ref{sec:kbe}--\ref{sec:ae} we overview KBE, GKBA or GKBA-WBL and second-order  AE equations, respectively, that we apply to these junctions. Results from these three methodologies---where those from KBE are numerically exact references and those from GKBA, GKBA-WBL and second-order AE are approximations that can be computed much faster---and their comparison are presented in Secs.~\ref{sec:resultspump}--\ref{sec:resultsinertia} for spin pumping, STT, and magnetic inertia examples, respectively. We conclude in Sec.~\ref{sec:conclusions}.

\section{Models and Methods}\label{sec:mm}

\subsection{Hamiltonian Models}\label{sec:hamiltonians}

The two-terminal junctions in Fig.~\ref{fig:fig0} consist of a two site C region sandwiched by the left (L) and the right (R) semi-infinite normal metal (NM) ideal leads. All three regions are modeled as one-dimensional (1D) TB chains. The two sites $(i=1,2)$ of the central region also host LMMs, modeled as classical vectors $\mathbf{M}_i$ of fixed unit length. The electronic TB Hamiltonian of each of the three junctions is given by 
\begin{eqnarray}\label{eq: total_hamil}
    \hat H(t) = \hat H_{C}(t)   +\sum_{p=L,R}  \hat H_p +\left (\hat{H}_{Cp} + \mathrm{H.c.}\right ).
\end{eqnarray}
Here, $\hat H_C(t) $ describes the C region 
	\begin{equation}\label{eq:hamil}
		\hat{H}_C(t)  =    - J_{sd}\sum_i\hat{c}_i^\dagger \hat{\bm \sigma} \cdot \mathbf{M}_i(t) \hat{c}_i- \gamma \sum_{\langle ij \rangle}  \big(\hat{c}_{i}^{\dagger} \hat{c}_{j} + \rm{H.c} \big),   
	\end{equation}
    which is time dependent due to the dynamics of LMMs $\mathbf{M}_i(t)$ that interact with the spin operator $\mathbf{\hat \sigma}$ of conduction electrons via \textit{sd} exchange coupling~\cite{Ralph2008,Cooper1967,Tsvelik2017}  $J_{sd}=0.1\gamma$.   The Hamiltonian of the lead $(p=L,R)$  is given by  
\begin{eqnarray}\label{eq: hamil leads}
    \hat H_p  =  eV_p(t)\sum_i\hat{c}_{i}^{\dagger} \hat{c}_{i}- \gamma \sum_{\langle ij \rangle}  \big(\hat{c}_{i}^{\dagger} \hat{c}_{j} + \rm{H.c} \big), 
\end{eqnarray}
while the coupling between the central region and the lead is described by  
\begin{equation}\label{eq: coupling_bath}
        \hat H_{Cp } = - \gamma r_p(t) \sum_{i \in C ,j\in   p }  \hat{c}_{i}^{\dagger} \hat{c}_{j}.
\end{equation}
Here, the vector $\hat c_i = (\hat c_{i \uparrow}, \hat c_{i\downarrow})$  contains operators that annihilate  an electron at site $i$ and with spin $\sigma = \uparrow, \downarrow$; similarly, $c^{\dagger}_{i\sigma}$ creates an electron in the same state; the nearest-neighbor (NN) hopping  is denoted by $\gamma$, and it sets the unit of energy; $V_p(t)$ is the time-dependent voltage  applied to lead $p$; and $r_p(t)$ is the switch-on function for attaching the lead $p$ to the $C$ region.

% Keeping in mind the issue of different stationary states~\cite{Dhar2006,Stefanucci2007,Khosravi2008,Stefanucci2008}, we use the same initial-state preparation for the three setups shown in Fig.~\ref{fig: fig0}.
We employ a  protocol~\cite{Ridley2022} in which the $C$ region is initially unoccupied and is then slowly connected to the leads using the ramp function  
\begin{equation}\label{eq:ramp}
r(t;\tau_{Cp})=\Theta(t)\left[\sin^2\!\left(\frac{\pi t}{2\tau_{ Cp}}\right)\Theta(\tau_{ Cp}-t)+\Theta(t-\tau_{ Cp})\right], 
\end{equation}
where $\tau_{Cp}$ is time to reach full connection and $\Theta(t)$ is the step function; and we denote $r_p(t)\equiv r(t;\tau_{Cp})$ when referring to the ramp function of lead $p$.  The LMMs are kept frozen until all transient  currents die away and the electronic subsystem reaches its Gibbs density matrix as the thermal equilibrium state. After that, we either initiate the  dynamics of LMMs due to external magnetic fields [Figs.~\ref{fig:fig0}(a) and \ref{fig:fig0}(c)] or pulsed bias voltage  $V_b(t)$ [Figs.~\ref{fig:fig0}(b)].

For the junction in Fig.~\ref{fig:fig0}(a), we set LMMs into steady precession, \mbox{ $\mathbf{M}_i(t) = \big (\sin \theta_0 \cos(\omega_0 t),\sin \theta_0 \sin(\omega_0 t),\cos \theta_0\big )$}, with frequency $\omega_0$ and cone angle $\theta_0$. In realistic experimental devices, such a  situation is created by, e.g., magnet absorbing microwave radiation under the ferromagnetic resonance condition~\cite{Ando2014a}. The spin current $I^{S_{\alpha}}_p = I^{\uparrow}_p - I^{\downarrow}_p$, where $I^{\uparrow}_p$ and $I^{\downarrow}_p$ are spin-resolved charge currents transporting  spin pointing in the $\alpha$-direction, which will be pumped by the dynamics of LMMs, is obtained from  
\begin{equation}\label{eq:spincurrent}
    I_p^{S_{\alpha}}(t) =   \mathrm{Tr}_\mathrm{spin} [ {\bm \rho}(t){\hat   I^{S_{\alpha}}_p} ].
\end{equation}
where $\hat{I}^{S_{\alpha}}_p$ is the spin current operator. 

\subsection{Kadanoff-Baym equations as a numerically exact reference for benchmarking}\label{sec:kbe}

The NEGF, or Schwinger-Keldysh-Kadanoff-Baym, formalism operates with four GFs~\cite{Stefanucci2025} out of which only two are independent. In several NEGF-based packages, such as {\tt NESSi}~\cite{Schueler2020} and {\tt CHEERS}~\cite{Pavlyukh2023}, the retarded $\mathbf{ G}^R(t,t')$ and the lesser $ \mathbf{G}^<(t,t')$ GFs are chosen as independent to take advantage of cheaper history integrals. Other choice is employed in packages like {\tt H-NESSi}~\cite{Blommel2026}, where the lesser and the greater $\mathbf{G}^>(t,t')$ GFs are used to achieve higher order numerical accuracy. For example, the equations of motion for the retarded and the lesser GFs are given by
\begin{subequations}\label{eq:KBE}
\begin{align} 
 \big[i\partial_t - \mathbf{ H}_C(t)\big ]\mathbf{G}^r(t,t') &= \mathbf{1}_C\delta(t,t') + \big (\mathbf{\Sigma}^r\circ \mathbf{ G}^r\big ) (t,t')\label{eq:dyson_gr.},
 \\ \mathbf{G}^<(t,t') &=  
 % \hat G^R(t,0)\hat G_{0}^{<}(0,0)\hat G^A(0,t') \nonumber
  \big (\mathbf{G}^r\circ\mathbf{ \Sigma}^{<}\circ \mathbf{ G}^a\big )(t,t')  \label{eq:dyson_gl.},
\end{align}
\end{subequations}
where the notation for the time convolution \mbox{$( \mathbf{A}\circ \mathbf{B})(t,t')$} is
\begin{equation}\label{eq:convolution}
(\mathbf{ A}\circ \mathbf{ B})(t,t') = \int^{\infty}_{t_0} d\bar  t \mathbf{ A}(t,\bar t )\circ \mathbf{ B}(\bar t, t').
\end{equation}
Note that $\mathbf{G}^r(t,t')$ the spectral properties of the system, while $\mathbf{G}^<(t,t')$ encodes how the spectrum is occupied out of equilibrium.  In particular, the time-evolved $\mathbf{G}^<(t,t')$ yields directly the nonequilibrium density matrix 
\begin{eqnarray}\label{eq:rho_from_lesser}
 {\bm \rho}(t) = -i \mathbf{ G}^<(t,t')\Big|_{t=t'}, 
\end{eqnarray}
which we need for describing phenomena in Fig.~\ref{fig:fig0} via Eqs.~\eqref{eq:sden} and ~\eqref{eq:spincurrent}. Throughout the paper, bold symbol $\mathbf{A}$ denotes matrix composed of elements $A_{i\sigma,j\sigma'}$ in the one-electron Hilbert space spanned by orbitals and spin of the central regions in Fig~\ref{fig:fig0}, so that, e.g.,  the $\mathbf{H}_C$ is the matrix  representation of the Hamiltonian in Eq.~\eqref{eq:hamil} of the size size $4 \times 4$. In addition,  $\mathbf{1}_C$ is the unit matrix multiplying the delta function $\delta(t,t')$;  $\mathbf{\Sigma}^R(t,t')$ and $\mathbf{\Sigma}^<(t,t')$  are the retarded and lesser, respectively, self-energies~\cite{Stefanucci2025} due to integrated-out semi-infinite NM leads; and we use $\hbar=1$  for simplicity. Note that,  in principle, $\mathbf{\Sigma}^<(t,t')$ should include an additional term $\delta(t-t_0)\mathbf{g}^{<}_0(t,t')\delta(t'-t_0)$, beyond those generated by the integrated out leads, due to different ways of connecting the leads to the C region~\cite{Stefanucci2004,Stefanucci2007} and possible quasibound states~\cite{Stefanucci2007}. Although they are usually neglected, we find that they can generate {\em significant} effect on magnetic inertia [Sec.~\ref{sec:resultsinertia}].

The total self-energies, \mbox{$\mathbf{\Sigma}^r(t,t') =  \mathbf{\Sigma}^{r}_L(t,t') + \mathbf{\Sigma}^r_R(t,t')$} and \mbox{$\mathbf{\Sigma}^<(t,t') =  \mathbf{\Sigma}^{<}_L(t,t') + \mathbf{\Sigma}^<_R(t,t')$}, due to the $p= L,R$  leads can be expressed as 
\begin{subequations}\label{eq:allselfenergies}
\begin{align}
\boldsymbol{\Sigma}^{<}_p(t,t')
&=
\!i\,
e^{-i\phi_p(t,t')}
\!\!\int\!\! \frac{dE}{2\pi}\,
\!f_p(E)\,
\boldsymbol{\Gamma}_p(E)\,
e^{-iE(t-t')},
\label{eq:self_lead_less}
\\
\boldsymbol{\Sigma}^{>}_p(t,t')
&=\!
-ie^{-i\phi_p(t,t')}
\!\!\int\!\! \frac{dE}{2\pi}\,
\!\bar f_p(E)\,
\boldsymbol{\Gamma}_p(E)\,
e^{-iE(t-t')},
\label{eq:self_lead_greater}
\\
\boldsymbol{\Sigma}^{r}_p(t,t')
&=
-i\Theta(t-t')
\left[
\boldsymbol{\Sigma}^{>}_p(t,t')
-
\boldsymbol{\Sigma}^{<}_p(t,t')
\right].
\label{eq:self_lead_retarded}
\end{align}
\end{subequations}
Here,  the Fermi function of the macroscopic reservoir kept at chemical potential $\mu_p$, into which lead $p$ terminates at infinity, is denoted by \mbox{$f_p(E)=\big [1+e^{(E-\mu_p)/k_BT}\big ]^{-1} $};  $\bar f_p(E) = 1 -f_p(E)$; and the level broadening  matrix~\cite{Popescu2016} is defined as \mbox{$\mathbf\Gamma_p(E) = -2\mathrm{Im}\big [\mathbf \Sigma^r_{p}(E)\big ]$} which can be computed analytically for square lattice leads and numerically exactly for other type of leads; $\phi_p(t,t') = \int^{t}_{t'}d\bar teV_p(\bar t) + i{\rm ln}\,[r_p(t)r_p(t')] $ accounts for the accumulated phase~\cite{Gaury2014,Tuovinen2023} due to time-dependent voltage $V_p(t)$ and the ramp function to reach full connection applied to lead $p$, respectively.

Since nonequilibrium Dyson Eqs.~\eqref{eq:KBE} are not suitable for numerical solving, because of involving the delta function $\delta(t,t')$, one first recasts them into the form of KBE for the lesser and greater GFs
\begin{subequations}\label{eq:kbeequation}
\begin{align}
    \big [i\partial_t-\mathbf{H}_C(t) \big ]\mathbf{G}^{<}(t,t') = \mathbf I^{<}(t,t'),\\
    \big [i\partial_t-\mathbf{H}_C(t) \big ]\mathbf{G}^{>}(t,t') = \mathbf I^{>}(t,t'),
\end{align}
\end{subequations}
where the collision integral $\mathbf{I}^{</>}(t,t')$ is given by 
\begin{align}\label{eq: collision intgrals (two-times)}
    \mathbf{I}^{<,>}(t,t') &= (\mathbf{\Sigma}^{<,>}\circ \mathbf{G}^a)(t,t') + (\mathbf{\Sigma}^r\circ \mathbf{G}^{<,>})(t,t').
    % I^{>}(t,t') &= (\Sigma^>\circ G^A)(t,t') + (\Sigma^R\circ G^{>})(t,t'),
\end{align}
%(G^<\circ\Sigma^A)(t,t') + (G^R\circ \Sigma^<)(t,t')
and $\mathbf{G}^a(t,t')= [\mathbf{G}^r(t',t)]^{\dagger} $ is the advanced GF. We solve KBE via an adaptive scheme of Ref.~\cite{Meirinhos2022}, which then provides a numerically exact benchmark for other methods explored, GKBA and second-order AE, as overviewed in Secs.~\ref{sec:gkba}--\ref{sec:ae}.

\subsection{Generalized Kadanoff-Baym ansatz}\label{sec:gkba}

The key idea of GKBA~\cite{Lipavsky1986} is to reduce the computational complexity of solving the full KBE [Eq.~\eqref{eq:kbeequation}] by reducing the numerical effort imposed by  the need to handle NEGF quantities depending on  two times. To achieve this goal of solving equations with only one discrete time variable, one first decouples the diagonal $t=t'$ from the off-diagonal $t\neq t'$ components of  the lesser GF via~\cite{Reeves2023,Reeves2024,Karlsson2021}
% , using and ansatz over the off diagonal components of the lesser GF
\begin{equation}\label{eq:GKBA}
       \mathbf{G}^<(t,t') \approx  \mathbf{\bm \rho}(t)\mathbf{G}^a(t,t')  -  \mathbf{G}^r(t,t')\mathbf{\bm \rho}(t').
\end{equation}
Note that ``$\approx$'' becomes ``$\equiv$'' for the case of closed noninteracting systems for which decomposition in Eq.~\eqref{eq:GKBA} is exact. Plugging in Eq.~\eqref{eq:GKBA} into KBE [Eq.~\eqref{eq:kbeequation}] does not lead to a closed equation for the nonequilibrium density matrix because   Eq.~\eqref{eq: collision intgrals (two-times)} requires $t\neq t'$ elements of the lesser and greater GFs. The GKBA solves this problem by employing~\cite{Reeves2023,Reeves2024,Karlsson2021}  either the mean-field approximation for $\mathbf{G}^r(t,t')$ or by predetermining the quasiparticle energy and damping of the retarded  GF, so that the lesser and greater GF become functionals of ${\bm \rho}(t)$.
%For open quantum systems, it has been proved~\cite{Balzer2023} that the use of the GKBA ansatz is equivalent to evolving the mixed GF while keeping the lead GF isolated
For our systems in Fig.~\ref{fig:fig0}, where the self-energies are generated by the NM leads while those~\cite{Karlsson2021} due to many-body interactions are absent, we can~\cite{Balzer2023} obtain an equation for  ${\bm \rho}(t)$ containing  quantities that depend on a single time variable
\begin{eqnarray} \label{eq:rho_C_embedding}
\lefteqn{i\partial_t{\bm\rho}(t)
=
\left[
\mathbf{H}_{C}(t),
{\bm \rho}(t)
\right]\nonumber} \\
&& \mbox{} - i\sum_p\bigg (
\mathbf{H}_{Cp}(t)\mathbf{G}_{pC}^{<}(t)
-
\mathbf{G}_{Cp}^{<}(t)\mathbf{H}_{pC}(t)
\bigg),
\end{eqnarray}
which is closed by using~\cite{Balzer2023}
\begin{eqnarray}\label{eq:closingeq}
i\partial_t\mathbf{G}_{pC}^{<}(t)
&=&
\mathbf{H}_{p}\mathbf{G}_{pC}^{<}(t)
-
\mathbf{G}_{pC}^{<}(t)\mathbf{H}_{C}(t)
\nonumber\\
&&
+i\mathbf{H}_{pC}(t)\boldsymbol{\rho}(t)
-
\mathbf{g}_{p}^{<}(t)\mathbf{H}_{pC}(t),
\\
i\partial_t\mathbf{g}_{p}^{<}(t)
&=&
\left[
\mathbf{H}_{p},
\mathbf{g}_{p}^{<}(t)
\right].\label{eq:glead}
\end{eqnarray}
Here, $\mathbf{G}_{pC}^{<}(t)$ is the matrix representation of the lesser GF, $ G^<_{pC, k j}(t)= -i\langle\hat c^{\dagger}_{k\in p}(t)\hat c_{j\in C}(t) \rangle $, in mixed real  and momentum space representation. Thus, this matrix is of the size $4\times N_k$, where $N_k$ is the number of $k$-points used in the discretization of the lead $p$. The quantity $\mathbf{g}_{p}^{<}(t)$ is the matrix representation of \mbox{$g_{p,k\in p}^{<}(t) = -i\langle\hat c^{\dagger}_{k}(t)\hat c_{k}(t) \rangle$}, as the lesser GF of the isolated lead $p$ and, therefore, it has dimension $N_k\times N_k$.

% , still it is possible to close the system of equation by finding 
% Using the ansatz to  The way in which this equat reconstruction of time-off diagonal elements re
% First, we should find an equation for the diagonal elements in time of the lesser Green function  $G^<(t,t^+)$, 
% \begin{align} \label{eq:reduced_density_matrix}
%     i\partial_t \hat G(t) &= [\hat H_s(t), \hat G (t)] + \hat I(t) + \hat I^{\dagger}(t),
% \end{align}
%where we have written the current operator in terms of the mixed GF, using \mbox{${\bm\Pi}_p(t) = \mathbf{G}_{Cp}^{<}(t)\mathbf{H}_{pC}(t)$}. Thus, the equation for of the mixed component can be closed using the GKBA ansatz

\subsection{GKBA-WBL methodology}\label{sec:gkbawbl}

Since GKBA is often (such as in the {\tt CHEERS} package~\cite{Pavlyukh2023}) further accelerated for two-terminal junctions by combining it with WBL approximation for the self-energies of  semi-infinite leads of the junction, we also benchmark this variation of GKBA methodology. In WBL approximation, one assumes that lead self-energies are energy independent, so \mbox{$\mathbf{\Gamma}_p(E) {\mapsto} \mathbf{\Gamma}_p(0)$}. Additionally, the Fermi function is represented using Pade decomposition
\begin{subequations}\label{eq:pade}
\begin{eqnarray}
    f(\omega) &=& \frac{1}{2} - \sum_n\eta_n \left(\frac{1}{\beta\omega+i\xi_n} +\frac{1}{\beta\omega-i\xi_n}    \right )
\end{eqnarray}
\end{subequations}
Then  Eq.~\eqref{eq:rho_C_embedding} simplifies into 
\begin{eqnarray}
\label{eq:WBL_rho}
i\partial_t\boldsymbol{\rho}(t)
&=&
\left[\mathbf H_C(t),\boldsymbol{\rho}(t)\right]
+
\sum_p \frac{i\,r_p^2(t)}{2}
\Big[
\boldsymbol{\Gamma}_p
-
\left\{
\boldsymbol{\Gamma}_p,\boldsymbol{\rho}(t)
\right\}\! \Big] \nonumber
\\
&+&
i\sum_{n,p}
r_p(t)\frac{\eta_n}{\beta}
\Big[
\boldsymbol{\Gamma}_p\mathbf{G}^{\mathrm{em}}_{n p}
+
\mathbf{ G}^{\mathrm{em}\dagger}_{n p}
\boldsymbol{\Gamma}_p
\Big],
\end{eqnarray}
which is closed by using 
\begin{eqnarray}
\label{eq:WBL_Gem}
i\partial_t
\mathbf{ G}_{n p}^{\mathrm{em}}(t)
&=& -r_p(t)\mathbf 1_C-\mathbf{G}_{n p}^{\mathrm{em}}(t)
\Big[\mathbf H_C(t)+\frac{i}{2}\sum_q s_q^2(t)\boldsymbol{\Gamma}_q
\nonumber\\
&-&\left( eV_p(t)-\frac{i\xi_n}{\beta}\right)\mathbf 1_C\Big],
\end{eqnarray}
where $\mathbf{G}^{\rm em}_{np}(t)$ is defined by~\cite{Tuovinen2023}
\begin{eqnarray} \label{eq:definition_GKBA_WBL_squeme}
    \mathbf{G}^{\rm em}_{np}(t) = \int d \bar t e^{-i\phi_p(t,\bar t) - (\xi_n/\beta)(t-\bar t)} \mathbf{G}^a(\bar t, t).
\end{eqnarray}
Note that the  usage of Pade decomposition [Eq.~\eqref{eq:pade}] makes it possible to access longer simulation times,  without restrictions imposed by the   finite number of  $k$-points (reflection effects start at times  $\sim N_k/\gamma$  which is the time taken by the electronic wave to go to the interface of the leads and come back) used to construct $\mathbf{G}_{pC}^{<}(t)$ and $\mathbf{g}_{p}^{<}(t)$ in Sec.~\ref{sec:gkba}.

\subsection{Second-order adiabatic expansion of NEGFs and thereby generated an extended LLG equation}\label{sec:ae}

Although AE can be conducted in several different frameworks---such as density matrix~\cite{Berry1993, Lenzig2025}, scattering matrix~\cite{Thomas2012} or NEGF~\cite{Bode2011,Bode2012,Deghi2024}---NEGF offers  the most versatile framework, as it can handle both electrons  interacting only with classical degrees of freedom~\cite{Bode2011,Bode2012,Lu2012,Dou2017a,Deghi2024} and electrons participating in additional quantum many-body interactions~\cite{Hopjan2018}. For example, the lowest-order object of NEGF theory describing electrons in two-terminal junctions without quantum many-body interactions is  adiabatic, or ``frozen-in-time,''  retarded GF~\cite{Salahuddin2006,Ellis2017}
\begin{equation}\label{eq:adiabaticgf}
\mathbf{G}^r(E,\bar T)=[E\mathbf{1}_C - \mathbf{H}_C(\bar T ) - \mathbf{\Sigma}^r_L(E) - \mathbf{\Sigma}^r_R(E)]^{-1}.
\end{equation}
This object is actually {\em identical} to retarded GF used in steady-state quantum transport~\cite{Datta1995,Waintal2024} but with  time $\bar T$ introduced and treated as a parameter only. Thus, Eq.~\eqref{eq:adiabaticgf} effectively assumes  that the velocity of the classical degree of freedom is zero (or infinitely slow). Nevertheless, the retarded GF in Eq.~\eqref{eq:adiabaticgf} has been frequently and self-consistently combined with LLG equation to model STT-induced dynamics of LMMs whose velocity is certainly not zero~\cite{,Xue2021}.  
Equation~\eqref{eq:adiabaticgf} is rigorously obtained as the lowest order term in the AE~\cite{Salahuddin2006,Ellis2017,Bode2011,Deghi2024} of  $\boldsymbol{\mathcal{G}}^r(E,\bar T)$ 
\begin{equation}\label{eq:wigner}
\boldsymbol{\mathcal{G}}^r(E,\bar T) = \int\!\! d\tau\, \mathbf{G}^r(t,t') e^{i E \tau/\hbar},
\end{equation}
which is the Wigner transform of $\mathbf{G}^r(t,t')$ using  \mbox{$\bar T=(t+t')/2$} and $\tau=t-t'$ as the so-called ``slow'' and ``fast'' time variables, respectively. The AE of \mbox{$\boldsymbol{\mathcal{G}}^r(E,\bar T)$} is given by 
\begin{eqnarray}\label{eq:gfexpansion}
\boldsymbol{\mathcal{G}}^r(E,\bar T)
&=& \mathbf{G}^r(E,\bar T)
+\sum_{\alpha,i}\boldsymbol{\Omega}^{r,\alpha}_{1,i}
\partial_t M^\alpha_i
\\
\lefteqn{\hspace{-6em}
+\Bigg\{
\sum_{\alpha,\alpha',i,j}
\boldsymbol{\Omega}^{\alpha\alpha',r}_{11,ij}
\partial_t M_i^\alpha\partial_t M_j^{\alpha'}
+\sum_{i,\alpha}
\boldsymbol{\Omega}^{\alpha,r}_{2,i}
\partial_t^2M^\alpha_i
\Bigg\} + \ldots \nonumber,
}
&& 
\end{eqnarray}
where the coefficients (see Ref.~\cite{Deghi2024} for explicit formulas)  $\bm\Omega_{1,i}^{r,\alpha}$, $\bm\Omega_{11,i}^{r,\alpha \alpha'}$ and  $\bm\Omega_{2,i}^{r,\alpha}$ require the knowledge of only $\mathbf{G}^r(E,\bar T)$, its derivatives over energy $E$ and time derivatives of classical degrees of freedom. In Eq.~\eqref{eq:gfexpansion} we use LMMs, $\mathbf{M}_i(t)$, as classical variables relevant to our study. Analogous expansion is also  performed~\cite{Bode2011,Deghi2024} for $\boldsymbol{\mathcal{G}}^<(E,\bar T)$ as the Wigner transform of $\mathbf{G}^<(t,t')$
\begin{eqnarray}\label{eq:gf_expansion_lesser}
\boldsymbol{\mathcal{G}}^<(E,\bar T) &=& \mathbf{G}^<(E,\bar T) + \sum_{\alpha,i}\boldsymbol{\Omega}^{<,\alpha}_{1,i} \partial_tM^{\alpha}_i \\
\lefteqn{\hspace{-6em}+ \Bigg\{\!\!\sum_{\alpha',\alpha,i,j} \boldsymbol{\Omega}^{\alpha \alpha',<}_{11,i j}\partial_tM_i^\alpha\partial_tM_j^{\alpha'} +\sum_{i,\alpha} \boldsymbol{\Omega}^{\alpha,<}_{2,i}\partial_t^2M^\alpha_i \Bigg \} + \ldots,} \nonumber
&&
\end{eqnarray}
where the zeroth-order term is obtained from  Eq.~(\ref{eq:adiabaticgf}) and its Hermitian conjugate $\mathbf{G}^A(E,\bar T) = [\mathbf{G}^R(E,\bar T)]^{\dagger}$ as the adiabatic lesser GF
\begin{equation}\label{eq: frozen_glesser}
\mathbf{G}^{<}(E,\bar T) = \mathbf{G}^{r}(E,\bar T)\mathbf{\Sigma}^{<}(E)\mathbf{G}^a(E,\bar T).
\end{equation}
Note that AE in Eqs.~\eqref{eq:gfexpansion} and \eqref{eq:gf_expansion_lesser}  were terminated to first-order terms, including $\partial_t M^{\alpha}_i$  in Refs.~\cite{Bode2011,Bode2012,Mahfouzi2016}; and to second-order terms including $\partial_t M_i^{\alpha} \partial_t M_j^{\alpha '}$   in Ref.~\cite{Deghi2024} or just including $\partial_t^2 M^{\alpha}_i$ in Ref.~\cite{Xu2023b}. 

The prefactor $\boldsymbol{\Omega}^{<,\alpha}_{1,i}$ of $\partial_t M^{\alpha}_i$ then provides NEGF-based expression for the Gilbert damping~\cite{Bode2012,Mahfouzi2016}; and the prefactor $\boldsymbol{\Omega}^{\nu,<}_{2,i}$ of $\partial_t^2 M^{\alpha}_i$ provides NEGF-based expression for the strength of  magnetic inertia. That is, by plugging in Eq.~\eqref{eq:gf_expansion_lesser}, via the STT term $\mathbf{T}_2$ [Eq.~\eqref{eq:stt}], into  into  the LLG Eq.~\eqref{eq:llg}, an effective equation for the classical dynamics of the LMMs is constructed with additional local and nonlocal Gilbert damping and magnetic  inertia terms
\begin{subequations}\label{eq:llgextended}
\begin{eqnarray}
	\partial_t \mathbf{M}_2 & =\! & -g_0 \mathbf{M}_2 \times \mathbf{B}^\mathrm{ext} + \alpha_G \mathbf{M}_2 \times \partial_t \mathbf{M}_2  \nonumber \\ 
    \mbox{} && +  \frac{g_0}{\mu_M} J_{sd} \langle \hat{\mathbf{ s}}_2\rangle \times \mathbf{M}_2, \\ 
    \langle \hat {\mathbf{ s}}_2\rangle & = & -i\int\!\! dE \, \mathrm{Tr}_\mathrm{spin}\, [\boldsymbol{\mathcal{G}}^<(E,\bar T){\bm \sigma}]. \label{eq:s2}
\end{eqnarray}
\end{subequations}
For example, Eq.~\eqref{eq:llgextended} includes  $\propto \mathbf{M}_1\times \partial_t\mathbf{M}_2$ and $\propto \mathbf{M}_1\times \partial^2_t\mathbf{M}_2$ for nonlocal damping and inertia, respectively. The prefactors of these terms are naturally {\em time-dependent}~\cite{Bajpai2019a,ReyesOsorio2025} rather than being static. Such terms are virtually never used in phenomenological LLG equations coded in classical micromagnetics~\cite{Moreels2026,Berkov2008} or ASD~\cite{Evans2014} packages. Since these coefficients depend only on Eqs.~\eqref{eq:gf_expansion_lesser}
and \eqref{eq: frozen_glesser}, they can be {\em precomputed} by using  many different configurations of $\mathbf{M}_i$. Thus constructed AE+LLG approach then offers a much faster alternative to either KBE+LLG or GKBA+LLG computational modeling of junctions in Fig.~\ref{fig:fig0}. Note that in extended LLG Eq.~\eqref{eq:llgextended}, we retain the standard local Gilbert damping with a static $\alpha_G$ parameter, which is often computed from first-principles~\cite{Guimaraes2019,Starikov2010} and related to intrinsic properties of a material like spin-orbit (SO) coupling or magnetic disorder
scattering~\cite{Starikov2010}. On the other hand, additional time-dependent local and nonlocal damping terms generated in Eq.~\eqref{eq:llgextended} are the consequence of electronic spin being always somewhat behind~\cite{Sayad2015} the motion of LMMs, as electron spin dynamics is never~\cite{Petrovic2021} infinitely fast regarding LMM dynamics, which then generates a nonzero STT term due to noncollinearity of $\langle \hat{\mathbf{ s}}_2\rangle(t)$ and $\mathbf{M}_2(t)$. Thus, such term does not require either SO coupling or magnetic disorder
scattering governing~\cite{Guimaraes2019, Starikov2010} $\alpha_G$. 

\section{Results and Discussion}\label{sec:results}

\subsection{Benchmarking GKBA and second-order AE vs. full KBE using   the spin pumping phenomenon}\label{sec:resultspump}

\begin{figure}
    \centering
    \includegraphics[width=1.0\linewidth]{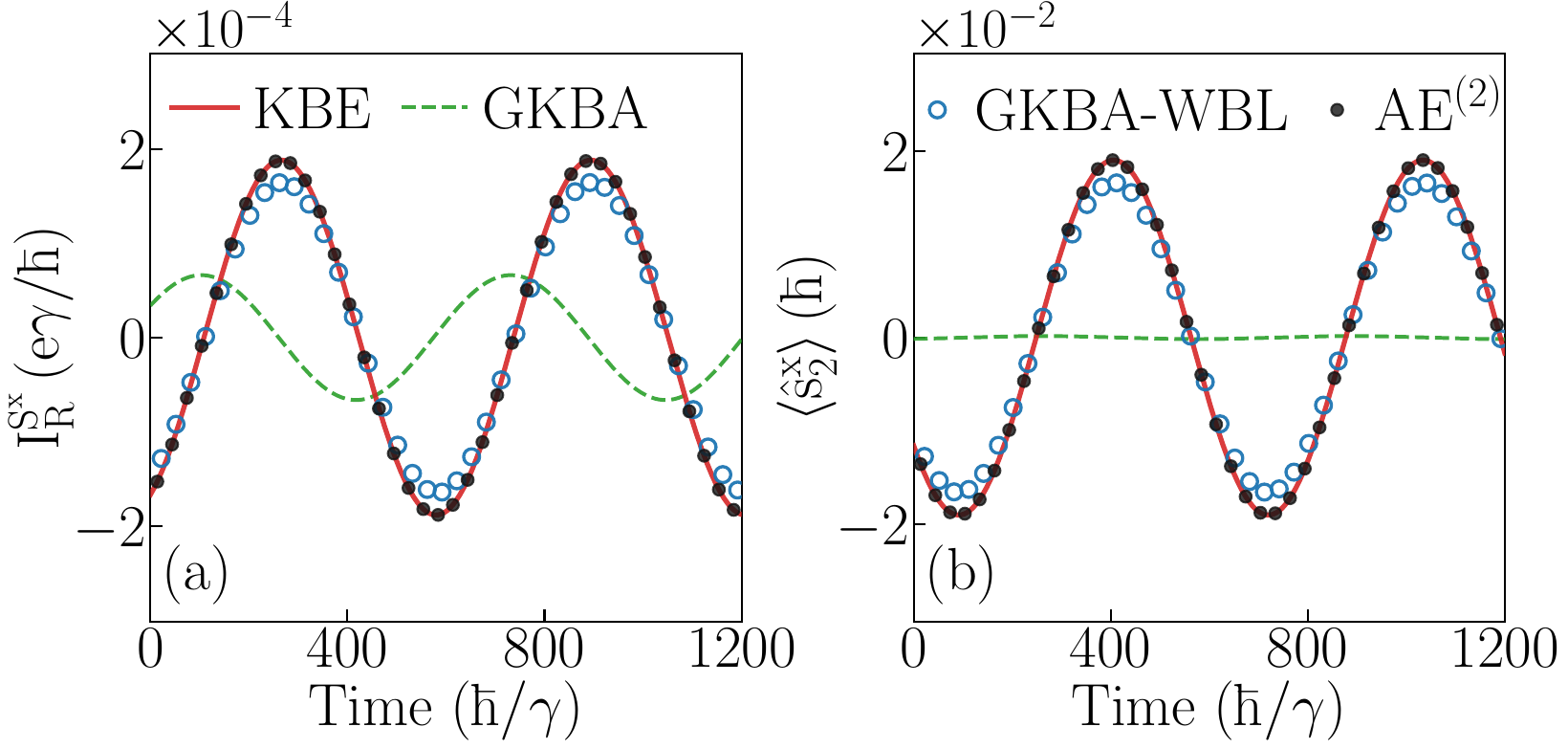}
    \caption{Time dependence of (a) spin current $I^{S_x}_R$ in the $R$ lead and (b) electronic spin density at the  site $i=2$, $\langle\hat{s}^{x}_{2}\rangle$,  in the spin pumping setup of Fig.~\ref{fig:fig0}(a)  where  two LMMs are steadily precessing in phase around  the  $z$-axis  with a cone angle $\theta_0 = 15^\circ$ and the precession frequency $\hbar\omega_0 = 0.01 \gamma $. Both panels employ $J_{sd} = 0.2 \gamma$ and no bias voltage is applied between the leads. }
    \label{fig:fig2}
\end{figure}

In time-dependent quantum transport calculations performed in this study, pumping of transient spin currents (see Fig.~6(a) in Ref.~\cite{Petrovic2018}) will be initiated at the time $t= -20\ \hbar/\gamma$ at which two LMMs are brought into steady precession [Fig.~\ref{fig:fig0}(a)]. After such transients die away, we obtain perfectly periodic $I_R^{S^x}(t)$ [red curve  in Fig.~\ref{fig:fig2}(a)] and constant $I_R^{S^z}$ [not explicitly shown in Fig.~\ref{fig:fig2}(a)]. Such periodicity or steadiness of pumped spin currents~\cite{Tserkovnyak2005} is ensured by the presence of semi-infinite NM leads~\cite{VarelaManjarres2023}, and reproduced by both GKBA and second-order AE schemes. However, GKBA fails to match the KBE result for spin current [Fig.~\ref{fig:fig2}(a)] and it performs even worse for nonequilibrium spin density [Fig.~\ref{fig:fig2}(b)]. Conversely, second-order AE matches both KBE-computed quantities perfectly in Fig.~\ref{fig:fig2}. Nevertheless, GKBA-WBL comes very close to the KBE result in Fig.~\ref{fig:fig2}. Note that KBE results in Fig.~\ref{fig:fig2} are both numerically and    analytically {\em exact} as they follow expressions from Ref.~\cite{Chen2009}.

Note that spin pumping by precessing magnetization is considered to be ``adiabatic'' because the pumped spin current is proportional~\cite{Tserkovnyak2005,Chen2009,Citro2023} to the frequency of magnetization precession, $I^{S^{\alpha}}_p \propto \omega$.  However, ``adiabatic'' terminology in quantum physics generally means that a time-dependent quantum system remains in its lowest energy state if closed, or it is described by objects like an  adiabatic density matrix~\cite{Berry1993,Bajpai2020} or adiabatic NEGFs in Eqs.~\eqref{eq:adiabaticgf} and \eqref{eq: frozen_glesser} when opened. From Fig.~\ref{fig:fig2} we learn that spin pumping is essentially a {\em nonadiabatic} phenomenon because the correct value of pumped current is recovered only by using NEGFs with second-order nonadiabatic corrections via Eqs.~\eqref{eq:gfexpansion} and \eqref{eq:gf_expansion_lesser}. The same conclusion about the {\em nonadiabatic} nature of spin pumping was also reached in Ref.~\cite{Tatara2019} using effective spin gauge fields. 

\subsection{Benchmarking GKBA-WBL and different orders of AE vs. full KBE using the STT phenomenon}\label{sec:resultsstt}

\begin{figure}[t]
    \centering
    \includegraphics[width=1.0\linewidth]{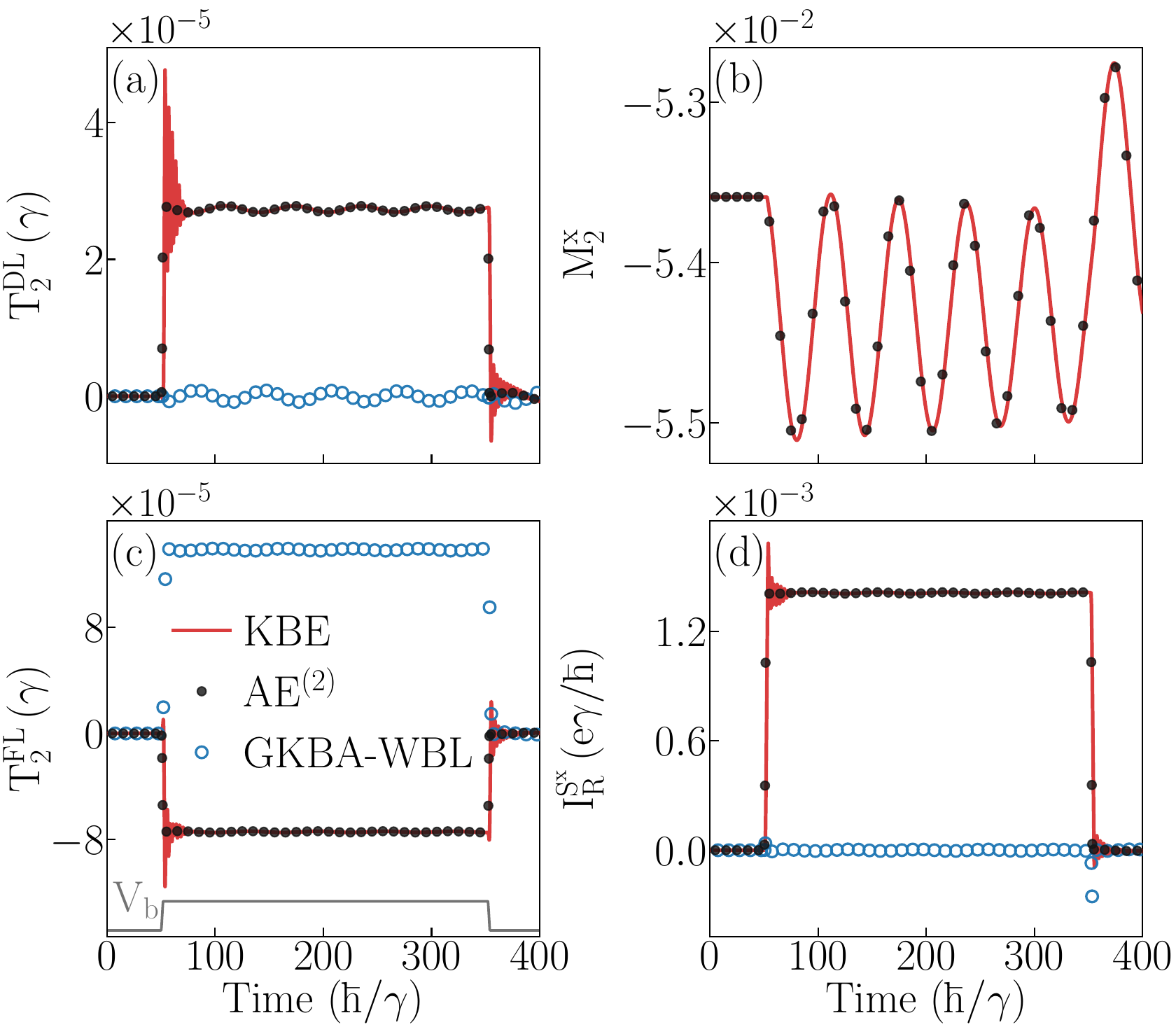}
    \caption{Time dependence of (a) DL and (c) FL components~\cite{Ralph2008,Petrovic2021} of the  STT vector [Eq.~\eqref{eq:FL-DL}] in the setup of Fig.~\ref{fig:fig0}(b). The STT acts on the free LMM $\mathbf{M}_2(t)$, stabilized near the $z$-axis by an applied magnetic field \mbox{$g_0\hbar B^{\rm ext} = 0.05 \gamma$}.  Panels (b) and (d) plot time dependence of ${M}^x_2$ and spin current $I^{S^x}_R$ pumped into the $R$ lead by the dynamics of the LMM $\mathbf{M}_2(t)$, respectively. All panels employ \mbox{$J_{sd} = 0.2 \gamma$} and $\alpha_G = 0$.  The voltage pulse [Eq.~\eqref{eq:ad_ramp}] injecting unpolarized charge current, which  drives STT  after being spin-polarized by $\mathbf{M}_1$ in Fig.~\ref{fig:fig0}(b), has amplitude  \mbox{$eV^{\rm \max}_b = 0.3\gamma$} and duration \mbox{$\tau_0=300 \hbar/\gamma$}, as illustrated by the gray line in panel (c).} 
    \label{fig:fig3}
\end{figure}

Since GKBA-WBL and AE passed the benchmark of Sec.~\ref{sec:resultspump}, we employ them for the next benchmark on the STT phenomenon studied using the junction in Fig.~\ref{fig:fig0}(b). For this phenomenon, the relevant quantities are the STT vector $\mathbf{T}_2$ on the second LMM in Fig.~\ref{fig:fig0}(b), thereby initiated the dynamics of $\mathbf{M}_2(t)$ the spin currents $I_p^{S^x}(t)$ that will be pumped into the NM leads. As usual in the spintronic literature, we split the  STT vector into the damping-like (DL) and field-like (FL) components 
\begin{equation}\label{eq:FL-DL}
    \mathbf{T}_2 = T^{\rm FL}_2 (\mathbf{M}_1\times\mathbf{M}_2) + T^{\rm DL}_2 \mathbf{M}_2\times(\mathbf{M}_1\times\mathbf{M}_2),
\end{equation}
whose time dependence in response to the voltage pulse 
\begin{equation}\label{eq:ad_ramp}
    V_b(t)=V^{\rm max}_b r (t;\tau_{\rm on})r (2\tau_{\rm on}+\tau_0-t;\tau_{\rm on}),
\end{equation}
is plotted in Figs.~\ref{fig:fig3}(a) and \ref{fig:fig3}(c). Here, $V^{\rm max}_b$ is the amplitude of the pulse and $\tau_{\rm on}$ is the time to turn it on or off and $\tau_0$ is the duration of the pulse plateau. The ensuing  $M_2^x(t)$  and $I_R^{S^x}(t)$  are plotted in Figs.~\ref{fig:fig3}(b) and \ref{fig:fig3}(d).  

\begin{figure}
    \centering
    \includegraphics[width=1.03\linewidth]{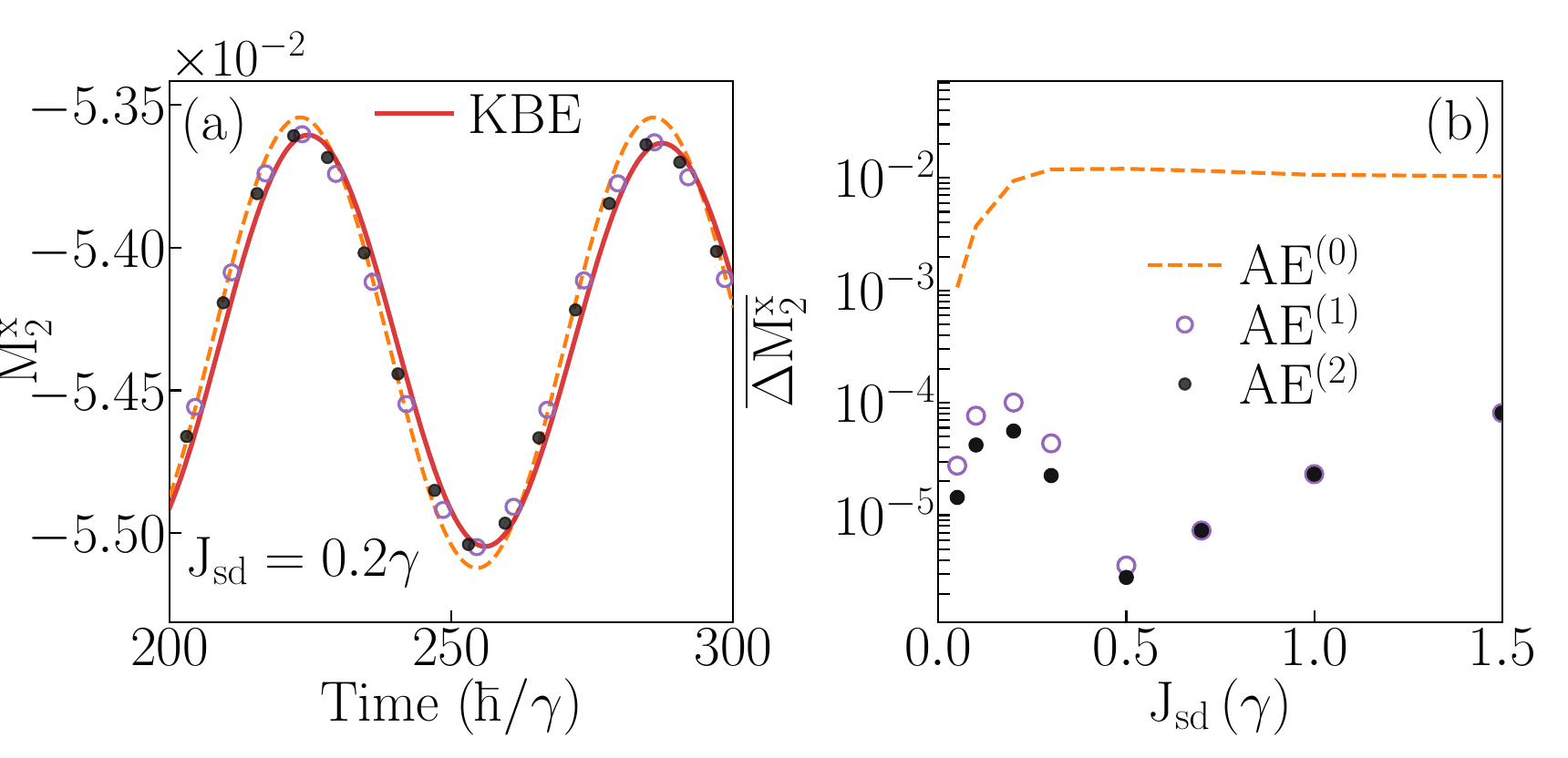}
    \caption{(a) Time dependence of $M_2^x$ from Fig.~\ref{fig:fig3}(b) within the interval $t \in [200,300]\hbar/\gamma$; and (b) error \mbox{$\overline{\Delta M_2^x} = M_2^x - M_2^{{x},\mathrm{KBE}}$}, averaged over the same  interval, as a function of $J_{sd}$. Here, $M_2^{x,\mathrm{KBE}}$ is red line from panel (a) as a {\em numerically exact} benchmark. The other three calculations are performed by using the first (AE$^{(0)}$ as zeroth-order AE~\cite{Ellis2017,Salahuddin2006}), second (AE$^{(1)}$ as first-order AE~\cite{Bode2011,Bode2012}) or  all three (AE$^{(2)}$ as second-order AE~\cite{Deghi2024}) terms on the right-hand side (RHS) of Eq.~\eqref{eq:gf_expansion_lesser} plugged into the LLG Eq.~\eqref{eq:llgextended}.}
    \label{fig:fig4}
\end{figure}

We see that GKBA-WBL fails completely for all four quantities in Fig.~\ref{fig:fig3} [note that we do not show $M^x_2(t)$ for GKBA-WBL in Fig.~\ref{fig:fig3}(d) due to being of the order $10^{-3}$]. On the other hand, second-order AE continues to perform well, except when bias voltage is changing too fast---compare fast oscillations of KBE-computed red curves  in Figs.~\ref{fig:fig3}(a)--(c) with second-order AE-computed black dots at times when the pulse is switched on and switched off. Figure~\ref{fig:fig4} clarifies that the first nonadiabatic correction in Eq.~\eqref{eq:gf_expansion_lesser} is crucial for this good performance of AE, while the second-order one offers negligible improvement. Note that in Fig.~\ref{fig:fig3} we denote usage of:  $\boldsymbol{\mathcal{G}}^<(E,\bar T) \approx \mathbf{G}^<(E,\bar T)$  as AE$^{(0)}$ or zeroth-order AE~\cite{Ellis2017,Salahuddin2006};  $\boldsymbol{\mathcal{G}}^<(E,\bar T) \approx \mathbf{G}^<(E,\bar T) + \sum_{\nu,i}\boldsymbol{\Omega}^{<,\nu}_{1,i} \partial_tM^{\nu}_i$ as AE$^{(1)}$ or first-order AE~\cite{Bode2011,Bode2012}; and full expression from Eq.~\eqref{eq:gf_expansion_lesser} as AE$^{(2)}$ or second-order AE~\cite{Deghi2024}.

\subsection{Benchmarking different orders of AE vs. full KBE using magnetic inertia phenomenon}\label{sec:resultsinertia}

Since  AE passed the benchmarks of both  Secs.~\ref{sec:resultspump} and \ref{sec:resultsstt}, we employ it for the final benchmark using the  magnetic inertia phenomenon exhibited by the junction in Fig.~\ref{fig:fig0}(c). In this junction, with no bias voltage applied, we apply an external magnetic field  that initiates the dynamics of two LMMs. The quantum subsystem of conduction electrons responds to their motion by going into a time-dependent nonequilibrium state described by $\mathbf{G}^r(t,t')$ and $\mathbf{G}^<(t,t')$ obtained by solving full KBE {\em exactly}; or described {\em approximately} and in computationally much less expensive fashion by their AE, $\boldsymbol{\mathcal{G}}^r(E,\bar T)$ [Eq.~\eqref{eq:gfexpansion}] and $\boldsymbol{\mathcal{G}}^<(E,\bar T)$ [Eq.~\eqref{eq: frozen_glesser}], respectively. The second-order AE of NEGFs then produces extended LLG Eq.~\eqref{eq:llgextended} containing an inertial term $\propto \mathbf{M}_i \times \partial_t^2\mathbf{M}_i$ that is also employed in the so-called~\cite{Mondal2023,Neeraj2020,De2025,Hartmann2025,Hartmann2026} inertial LLG (iLLG) equation. However, unlike the phenomenological iLLG equation, our extended  LLG Eq.~\eqref{eq:llgextended} also contains nonlocal terms $\propto \mathbf{M}_i \times \partial^2_t\mathbf{M}_j$ and $\propto \mathbf{M}_i \times \partial_t \mathbf{M}_j \partial_t \mathbf{M}_k$.  

\begin{figure}
    \centering
    \includegraphics[width=1.03\linewidth]{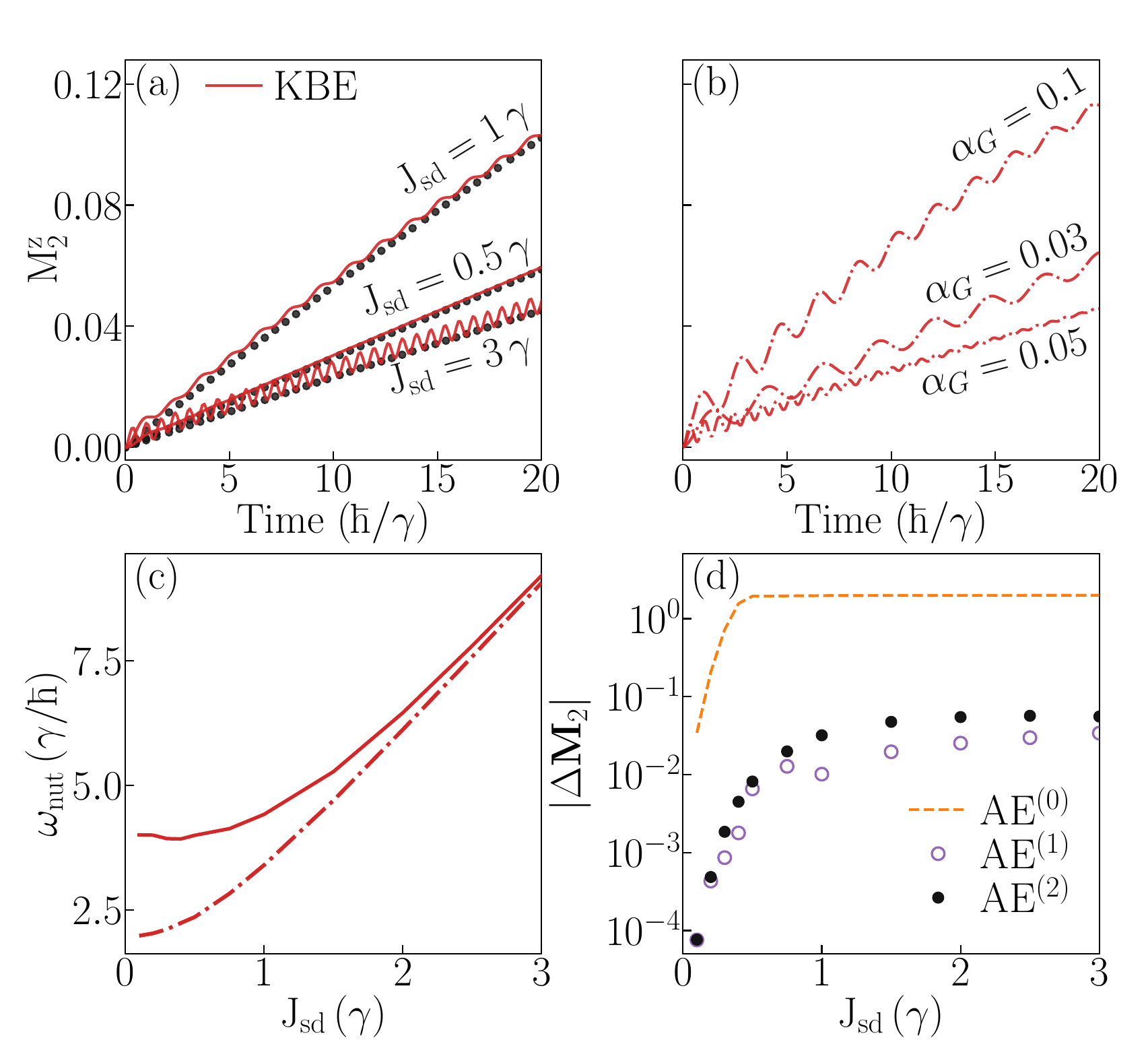}
    \caption{Time dependence of the  $z$-component of LMM vector $M^z_2$ in the magnetic inertia setup of Fig.~\ref{fig:fig0}(c) for (a) NM leads fully connected and (b) disconnected. Both solid red and dashed-dot red curves in  panels (a) and (b) are computed exactly by full KBE.  Panel (c) plots $J_{sd}$ dependence of  frequency $\omega_{\rm nut}$ of nutational oscillations~\cite{Mondal2023,Svingen2025} from panels (a) and (b) that are observable hallmark~\cite{Mondal2023,De2025,Neeraj2020} of magnetic inertia. Panel (d) plots $J_{sd}$ dependence of the error \mbox{$|\Delta \mathbf{M}_2| = |\mathbf{M}_2 - \mathbf{M}_2^{\mathrm{KBE}}|$} computed at selected times when such error is maximized.  The dynamics of two LMMs in Fig.~\ref{fig:fig0}(c) is initiated by applying an external magnetic field $g_0\hbar B^{\rm ext} = 0.1 \gamma$.}
    \label{fig:fig5}
\end{figure}

Surprisingly, despite having all terms of the iLLG equation, our extended LLG Eq.~\eqref{eq:llgextended} does not capture fast nutational oscillation  superimposed on regular damped  trajectories of LMMs in Fig.~\ref{fig:fig5}(a). These fast nutational oscillations are obtained from the full KBE-computed reference results in which  $\boldsymbol{\mathcal{G}}^<(E,\bar T)$ is replaced by $\mathbf{G}^<(t,t)$ in Eq.~\eqref{eq:s2}. In other words, KBE-computed numerically exact reference result is equivalent to using AE of $\boldsymbol{\mathcal{G}}^<(E,\bar T)$ in Eq.~\eqref{eq:gf_expansion_lesser} to {\em all} orders. Another surprise is that second-order AE [AE$^{(2)}$ in Fig.~\ref{fig:fig5}(d)] performs worse than the first-order one [AE$^{(1)}$ in Fig.~\ref{fig:fig5}(d)]. 
Another way to interpret this surprising discrepancy is that the full KBE-generated LLG equation has a non-Markovian form~\cite{Bajpai2019a} which is not properly replicated by the second-order AE-generated LLG Eq.~\eqref{eq:llgextended}. Interestingly, very recent analyses~\cite{Hartmann2025,Hartmann2026} of experiments on THz-induced dynamics of magnetization in Co films also finds that neither standard nor inertial phenomenological LLG equations can describe their data, but a properly parameterized non-Markovian LLG equation can. 

Figure~\ref{fig:fig5}(b) shows that disconnecting NM leads maintains fast nutational oscillations of LMMs, as long as $\alpha_G$ of such closed system is increased as also observed in Ref.~\cite{Sayad2016}. Finally, frequency of nutational oscillations $\omega_\mathrm{nut}$ is given  in Fig.~\ref{fig:fig5}(c), becoming identical in junctions with and without NM leads as $J_{sd}$ is increased.

\section{Conclusions and Outlook}\label{sec:conclusions}

\subsection{Conclusions}\label{sec:conclude}

We summarize the main result of our benchmarking of GKBA, GKBA-WBL and second-order AE using spin pumping, spin-transfer torque, and magnetic inertia time-dependent effects in spintronics  as  follows: 

\begin{itemize}
    \item GKBA~\cite{Stefanucci2025,Balzer2023,Tuovinen2023} calculations fail [Fig.~\ref{fig:fig2}] to reproduce the exact benchmarks from full KBE calculations for  spin current  pumped by steadily precessing magnetization [Fig.~\ref{fig:fig0}(a)]. Surprisingly, GKBA-WBL methodology produces much better results than GKBA alone. However, GKBA-WBL fails completely [Fig.~\ref{fig:fig3}] to reproduce the STT vector and thereby induced magnetization dynamics. 
    
    \item Second-order AE~\cite{Deghi2024}  of NEGFs works remarkably well for the examples of spin pumping [Fig.~\ref{fig:fig2}] and STT  [Fig.~\ref{fig:fig3}]. This finding also confirms~\cite{Tatara2019} that spin pumping is essentially a {\em nonadiabatic} phenomenon [Sec.~\ref{sec:resultspump}].

    \item Despite including terms up to the second time-derivatives of LMMs into AE of NEGFs [Eq.~\eqref{eq:gf_expansion_lesser}]---which then inserts those derivatives into the LLG Eq.~\eqref{eq:llgextended} to mimic the structure of phenomenological iLLG equation~\cite{Mondal2023,Hartmann2025,Hartmann2026}, but with time-dependent coefficients and additional nonlocal terms--- second-order AE fails to describe fast nutational oscillations of local magnetization as the observational hallmark of magnetic inertia~\cite{Mondal2023,De2025,Neeraj2020}. These findings suggests the need for fully {\em non-Markovian}  LLG~\cite{Bajpai2019a,Hartmann2025,Hartmann2026,ReyesOsorio2025} to describe magnetization dynamics at high (THz) frequencies, but it does not affect our conclusion about the usefulness of second-order AE for magnetization dynamics at lower (GHz) frequencies encountered in spin pumping and STT phenomena.  
\end{itemize}

The failure of GKBA can be traced to the usage of simple,  when compared to those from many-body interactions~\cite{Stefanucci2025,Kalvova2024,Reeves2023}  but still nonlocal in time, self-energies of NM leads  [Eq.~\eqref{eq:allselfenergies}]. They increase the relevance of the off-diagonal~\cite{Reeves2023,Reeves2024} $t \neq t'$ elements of NEGF that are neglected by GKBA. In other words, NM leads make memory integrals of the full KBE important even in the absence of many-body interactions.  Additionally,  Eq.~\eqref{eq:glead}  shows that GKBA time-evolves lesser GF of NM leads independently from the  evolution of the $C$ region, thus violating conservation of spin and charge~\cite{Balzer2023,Kalkova2019}.  Note that even extended GKBA (eGKBA)~\cite{Balzer2023,Tuovinen2023,Hopjan2026} will not resolve these issues.  Nevertheless,  another very recently proposed~\cite{Pavlyukh2025} scheme iterating GKBA further (iGKBA) holds promise (yet to be demonstrated) to resolve the issues we found in time- and spin-dependent quantum transport, while maintaining linear scaling and being able to include~\cite{Pavlyukh2025a} many-body interactions within multiterminal junctions.

%--------------------------------------------------------
The failure of second-order AE for magnetic inertia can be traced to  dynamically excited bound states outside of the energy band  of NM leads due to the applied external magnetic field $\mathbf{B}^{\rm ext}(t)$.
% which is a manifestation of the localization of the electronic spin density $\langle\hat{\mathbf{s}}_i\rangle$ in the $C$ region. 
Consequently, a non-analytical resonance~\cite{Chakraborty2018} appears in the memory-kernel of the non-Markovian LLG equation~\cite{Hartmann2025,Hartmann2026}, which cannot be reproduced by AE. This is because AE is based on an expansion near zero frequency.

\subsection{Outlook}\label{sec:outlook}

In spintronics, substantial effort has been devoted over the past two decades to {\em first-principles}  calculations of STT~\cite{Wang2008b,Nikolic2018,Ellis2017} and its variants like SO torque~\cite{Nikolic2018,Belashchenko2019,Xue2021,Dolui2020}, as well as coupling of thus computed STT to the LLG equation~\cite{Ellis2017,Salahuddin2006,Dolui2020,Xue2021}. The coupling makes it possible to obtain the ensuing current-driven magnetization dynamics that is key ingredient of potential STT applications~\cite{Locatelli2014}. Many such calculations~\cite{Nikolic2018,Ellis2017,Belashchenko2019,Xue2021,Dolui2020} rely on the so-called NEGF+DFT~\cite{Brandbyge2002} framework, where a Hamiltonian from noncollinear density functional theory (DFT) is plugged into  adiabatic GFs [Eqs.~\eqref{eq:adiabaticgf} and ~\eqref{eq: frozen_glesser}]. So, they are already quite computationally expensive due to the demands of noncollinear DFT and $k$-point sampling~\cite{Wang2008b}, so that shift to fully time-dependent DFT calculations is a no-go. At the same time, usage~\cite{Ellis2017,Salahuddin2006} of only adiabatic GFs  can produce significant errors [Fig.~\ref{fig:fig4}] in the trajectories of current-driven magnetization dynamics. Our study demonstrates that computationally inexpensive second-order AE~\cite{Deghi2024} can provide quite an accurate description of all STT-related quantities [Figs.~\ref{fig:fig3} and ~\ref{fig:fig4}] while remaining easily interfaced with DFT calculations and amenable to 
precomputing all coefficients in Eqs.~\eqref{eq:gfexpansion} and ~\eqref{eq:gf_expansion_lesser} before time-dependent simulations of STT-driven magnetization dynamics.

\acknowledgments
This work was supported by the US Department of Energy (DOE) under Grant No. DE-SC0026068.

\bibliography{biblio}

\end{document}